\documentclass[sigconf]{acmart}
\AtBeginDocument{%
  }

\makeatletter
\def\@formatdoi#1{\begingroup\def\@tempa{#1}\ifx\@tempa\@empty\else\href{https://doi.org/#1}{https://doi.org/#1}\fi\endgroup}
\makeatother

\usepackage{subfigure}
\usepackage{tabularx}
\usepackage{listings}
\usepackage{tikz}
\usepackage{amsmath}
\usepackage{amsthm}
\usepackage{caption}
\usepackage{comment}
\usepackage{enumitem}
\usepackage{algorithm}
\usepackage{algpseudocode}
\usepackage{balance}
\usepackage{seqsplit}
\usepackage[a-2b]{pdfx}
\usepackage{float}

\makeatletter
    \newcommand*{\algrule}[1][\algorithmicindent]{\makebox[#1][l]{\hspace*{.5em}\thealgruleextra\vrule height \thealgruleheight depth \thealgruledepth}}%
\newcommand*{\thealgruleextra}{}
\newcommand*{\thealgruleheight}{.75\baselineskip}
\newcommand*{\thealgruledepth}{.25\baselineskip}

\newcount\ALG@printindent@tempcnta
\def\ALG@printindent{%
    \ifnum \theALG@nested>0
        \ifx\ALG@text\ALG@x@notext
        \else
            \unskip
            \addvspace{-1pt}
            \ALG@printindent@tempcnta=1
            \loop
                \algrule[\csname ALG@ind@\the\ALG@printindent@tempcnta\endcsname]%
                \advance \ALG@printindent@tempcnta 1
            \ifnum \ALG@printindent@tempcnta<\numexpr\theALG@nested+1\relax
            \repeat
        \fi
    \fi
    }%
\usepackage{etoolbox}
\patchcmd{\ALG@doentity}{\noindent\hskip\ALG@tlm}{\ALG@printindent}{}{\errmessage{failed to patch}}
\makeatother

\newbox\statebox
\newcommand{\myState}[1]{%
    \setbox\statebox=\vbox{#1}%
    \edef\thealgruleheight{\dimexpr \the\ht\statebox+1pt\relax}%
    \edef\thealgruledepth{\dimexpr \the\dp\statebox+1pt\relax}%
    \ifdim\thealgruleheight<.75\baselineskip
        \def\thealgruleheight{\dimexpr .75\baselineskip+1pt\relax}%
    \fi
    \ifdim\thealgruledepth<.25\baselineskip
        \def\thealgruledepth{\dimexpr .25\baselineskip+1pt\relax}%
    \fi
    \State #1%
    \def\thealgruleheight{\dimexpr .75\baselineskip+1pt\relax}%
    \def\thealgruledepth{\dimexpr .25\baselineskip+1pt\relax}%
}
\acmDOI{}
\acmISBN{}

\begin{document}

\algnewcommand\algorithmicswitch{\textbf{switch}}
\algnewcommand\algorithmiccase{\textbf{case}}
\algdef{SE}[SWITCH]{Switch}{EndSwitch}[1]{\algorithmicswitch\ #1\ \algorithmicdo}{\algorithmicend\ \algorithmicswitch}%
\algdef{SE}[CASE]{Case}{EndCase}[1]{\algorithmiccase\ #1}{\algorithmicend\ \algorithmiccase}%
\algtext*{EndSwitch}%
\algtext*{EndCase}%

\title{Breaking the Barriers of Database-Agnostic Transactions}
\author{Toshihiro Suzuki}
\affiliation{%
    \institution{Scalar, Inc.}
  \state{Tokyo}
  \country{Japan}
}
\authornote{Both authors contributed equally to this paper.}
\email{toshihiro.suzuki@scalar-labs.com}

\author{Hiroyuki Yamada}
\affiliation{%
    \institution{Scalar, Inc.}
  \state{Tokyo}
  \country{Japan}
}
\authornotemark[1]
\email{hiroyuki.yamada@scalar-labs.com}
\orcid{0000-0002-7137-0899}

\begin{abstract}
Federated transaction management has long been used as a method to virtually integrate multiple databases from a transactional perspective, ensuring consistency across the databases. Modern approaches manage transactions on top of a database abstraction to achieve database agnosticism; however, these approaches face several challenges. First, managing transactions on top of a database abstraction makes performance optimization difficult because the abstraction hides away the details of underlying databases, such as database-specific capabilities.
Additionally, it requires that application data and the associated transaction metadata be colocated in the same record to allow for efficient updates, necessitating a schema migration to run federated transactions on top of existing databases.
This paper introduces a new concept in such database abstraction called \textit{Atomicity Unit} (AU) to address these challenges.
AU enables federated transaction management to aggressively push down database operations by making use of the knowledge about the scope within which they can perform operations atomically, fully harnessing the performance of the databases. Moreover, AU enables efficient separation of transaction metadata from application data, allowing federated transactions to run on existing databases without requiring a schema migration or significant performance degradation. In this paper, we describe AU, how AU addresses the challenges, and its implementation within ScalarDB, an open-sourced database-agnostic federated transaction manager. We also present evaluation results demonstrating that ScalarDB with AU achieves significantly better performance and efficient metadata separation.
\end{abstract}

\maketitle

\section{Introduction}

Composing a system with multiple databases is a standard architecture in enterprises.
Decentralized data architectures such as microservices \cite{intro/microservices,intro/cerny,microservices/laigner} and data meshes \cite{microservices/newman,datamesh/goedegebuure,intro/datamesh} respect domain boundaries; thus, a system across multiple business domains typically has respective database instances.
One size does not fit all mantra \cite{intro/stonebraker} further accelerates the use of different kinds of databases in a system.
For example, each microservice chooses a database based on the service’s use case and the developers’ experiences.

Federated transactions, also known as multi-database transactions, have been widely used for managing transactions across multiple, possibly different kinds of, databases.
These have been explored since around the 1990s and progressed to meet new demands, such as supporting various query notations, providing all the functionalities of underlying databases, and providing distributed transactions across multiple databases that do not support the same transaction model.

Particularly, modern federated transaction managers have progressed to achieve database-agnostic transactions for wide applicability.
These approaches abstract databases and implement a transaction logic on top of the abstraction; thus, their transactions can naturally achieve transactions across multiple databases without depending on specific database implementations.

However, these approaches face several challenges.
First, managing transactions on top of a database abstraction makes performance optimization difficult because the abstraction hides away the details of underlying databases, such as database-specific capabilities, configurations, and optimizations.
As a result, for example, all read and write operations on the abstraction are treated as individual transactions to the underlying database, which results in significant performance overhead due to numerous transactions and a loss of intra-transaction optimization opportunities.
Additionally, these approaches require that application data and its corresponding transaction metadata be colocated in the same record to enable efficient, atomic updates, necessitating a schema migration to run federated transactions on top of existing databases.

This paper discusses a new concept in such database abstraction of database-agnostic federated transaction management called \textit{Atomicity Unit} (AU) to address these challenges.
AU extends the database abstraction by incorporating database-specific information, particularly the maximum scope (e.g., database, schema, table, or partition) within which a database can guarantee atomicity for read and write operations.
By introducing such simple information into the database abstraction and making each database adapter that implements the abstraction declare its atomicity unit, database-agnostic federated transaction managers can make their processing more efficient and more applicable.

In this paper, we propose two techniques based on AU: \textit{Atomicity Unit Pushdown (AUP)} and \textit{AU-Scoped Decoupling (AUD)}.
First, AUP enables a database-agnostic federated transaction manager to aggressively push down database operations by making use of the AU information of each database.
Specifically, the transaction manager with AUP can effectively group writes for each AU of the underlying databases.
As a result, each underlying database accepts grouped writes in a batch and writes them as a single transaction, reducing the number of writes to disk and fully harnessing the performance of the databases.
Second, AUD enables efficient separation of transaction metadata from application data, allowing federated transactions to run on existing databases without requiring a schema migration or significant performance degradation.
Specifically, AUD allows the transaction metadata to be separately managed from its application data in a database within the database's atomicity unit, so that writes to both data can be grouped in the database's transaction.
Although the number of reads is also doubled, AUD optimizes processing to minimize performance degradation.
It achieves this by further extending the database abstraction and leveraging the \textit{consistent readability} of a database, which allows for the consistent reading of multiple data items within an atomicity unit.

This paper makes the following contributions:
\begin{itemize}[leftmargin=*]
  \item We introduce \textit{atomicity unit} (AU), a simple yet effective concept that enriches the database abstraction of database-agnostic federated transaction managers with information about the scope within which each underlying database can guarantee atomicity.
  \item We propose \textit{atomicity unit pushdown} (AUP), a technique that groups write operations by their atomicity units to reduce the number of database transactions and harness database-specific optimizations, significantly improving transaction throughput.
  \item We propose \textit{AU-scoped decoupling} (AUD), a technique that enables efficient separation of transaction metadata from application data by leveraging the consistent readability and view joinability of underlying databases, allowing federated transactions to run on existing databases without schema migration.
  \item We implement both techniques in ScalarDB, an open-source database-agnostic federated transaction manager, and evaluate them using YCSB workloads, demonstrating that AUP approaches ideal performance and AUD limits the overhead of metadata decoupling to approximately 30\%.
\end{itemize}

The remainder of the paper is organized as follows.
Section~2 provides background on federated transactions and describes the challenges of database-agnostic approaches.
Section~3 defines atomicity unit and presents the two proposed techniques, AUP and AUD, along with their algorithms.
Section~4 evaluates the performance of both techniques.
Section~5 discusses related work, and Section~6 concludes the paper.

\section{Background and Motivation}
This section describes the approaches for executing federated transactions and the challenges of modern database-agnostic federated transactions.

\subsection{Federated Transactions}
Federated transactions, also known as multi-database transactions, are transactions that span multiple databases to provide global transactions with certain correctness guarantee, typically ACID.

Early research work \cite{intro/sheth,mdbs/breitbart,mdbs/mehrotra2,mdbs/geo,breitbart1988multidatabase,mdbs/elmagarmid,mdbs/mehrotra} and some recent work \cite{mdbs/tang} use the concurrency control mechanisms of the underlying databases.
Therefore, it cannot execute transactions across databases that do not have concurrency control mechanisms, such as NoSQL databases.
Similarly, early production systems \cite{related/tuxedo,related/seata,related/atomikos} based on the X/Open XA standard \cite{dt/xa} execute federated transactions only across XA-compliant databases, primarily traditional relational databases.
We call the transactions of these approaches database-dependent federated transactions.

\subsection{Database-agnostic Federated Transactions}

Modern approaches \cite{dey-cherrygarcia,yamada-scalardb,kraft-epoxy} achieve database-agnostic federated transactions by implementing a transaction protocol outside of databases, within applications or middleware that sits between applications and databases. This allows transactions to run across various types of databases, including NoSQL and modern NewSQL databases, which do not comply with the specifications used by database-dependent federated transactions.

These approaches typically define a database abstraction for database operations, such as reading and writing records, and implement transaction protocols on top of the abstraction to realize the database-agnostic property. Specifically, these manage transaction metadata, such as transaction logs and version timestamps, alongside application records, exploiting the metadata when reading and writing records instead of relying on database-internal logging mechanisms or data structures.

Let us take a closer look at one example, ScalarDB.
ScalarDB abstracts a database with operations and a data model.
It defines several operations, such as reading and writing operations.
Also, it abstracts a database as a multi-dimensional map, an extended key-value model similar to the Bigtable \cite{related/chang2} data model.
In the abstraction, a record (i.e., ScalarDB record) comprises partition keys, clustering keys, and a set of columns. The combination of partition and clustering keys form a primary key, and a primary key uniquely maps a record. Records with the same partition keys form a partition and are assumed to be sorted by clustering keys.
ScalarDB provides an adapter for each database which implements the abstraction.
For database-agnostic transactions, ScalarDB treats each record of each database as a small separate database and performs a variant of the two-phase commit (2PC) on multiple records.
To regard each record as a database, ScalarDB separately manages write-ahead logging (WAL) information for each record, called distributed WAL (DWAL).
Specifically, it adds transaction metadata corresponding to WAL information to a record besides the columns that an application manages.
The transaction metadata for a record includes, for instance, the ID (TxID) of the transaction that last updated the record, the record's version number (Version), the record's state (TxState) (i.e., COMMITTED or PREPARED), timestamps, and a before image consisting of the previous version of the application data and its metadata.
It also manages transaction states separately from the application records in a table called a coordinator table.
The coordinator table determines and maintains transaction states as a single source of truth.
ScalarDB employs optimistic concurrency control (OCC) as its concurrency control protocol.
More details about the transaction protocol can be found in the ScalarDB paper \cite{yamada-scalardb}.

\subsection{Challenges of Database-agnostic Federated Transactions} \label{subsec:challenges}

Although modern approaches enable database-agnostic federated transactions, these face several challenges.
First, these approaches implement transaction protocols on top of a database abstraction. This makes it technically difficult to optimize transaction performance. The abstraction hides important details, such as database-specific capabilities, configurations, and optimizations, preventing transaction protocols from leveraging those pieces of information for performance enhancements.
For example, all read and write operations on the abstraction are treated as individual transactions to the underlying database. This results in significant performance overhead due to numerous transactions and a loss of intra-transaction optimization opportunities because of the separation of transactions.

Second, these approaches require that application data and the associated transaction metadata be colocated in the same record to allow for efficient updates. This necessitates a schema migration for running federated transactions on existing databases. While it is possible to manage these elements separately in different locations, such as in different tables, doing so effectively doubles the number of read and write operations to the underlying database, which can severely impact performance.

This paper introduces a new concept in such database abstraction called \textit{Atomicity Unit} to address these challenges without losing the benefits of the approaches.

\section{Atomicity Unit} \label{sec:au}

\textit{Atomicity unit} (AU) is the maximum scope (e.g., database, schema, table, or partition) within which a database can guarantee atomicity for read and write operations.
For example, 
the AU of traditional relational database management systems (RDBMS), such as PostgreSQL, MySQL, and Oracle Database, as well as some NoSQL databases like DynamoDB, and NewSQL databases, such as CockroachDB and YugabyteDB is \textit{database}. In contrast, the AU of certain NoSQL databases, such as Cosmos DB and Cassandra, is \textit{partition}.

By introducing such a simple concept into database abstraction, database-agnostic transaction managers built on top of it can utilize database-specific capabilities without breaking the abstraction, enabling their processing to be more efficient and more applicable.
In this section, we describe two techniques based on AU: \textit{atomicity unit pushdown} and \textit{AU-scoped decoupling}.

\subsection{Atomicity Unit Pushdown} \label{subsec:aup}

This section explains the overview of \textit{atomicity unit pushdown} (AUP) and its algorithm.
To illustrate how AUP works, we will use ScalarDB as an example of a database-agnostic federated transaction manager; however, this approach is applicable in general.

\begin{algorithm}[htbp]
\caption{Atomicity Unit Pushdown in the Transaction Manager Layer.}
  \label{alg:aup}
  \begin{algorithmic}[1]
  \Function {commit}{TxID, readSet, writeSet} \label{alg:aup:commit}
   \State groups $\gets$ \Call{groupByAtomicityUnit}{writeSet} \label{alg:aup:groupby-aup}
      \If {groups.size == 1 \textbf{and} !\Call{ensureSerializable}{}} \label{alg:aup:one-phase-begin}
      \State records $\gets$ \Call{createCommittedRecords}{groups[0]}
      \State \Call{atomicWrite}{records} \Comment{one-phase commit}
      \State return
    \EndIf \label{alg:aup:one-phase-end}
    \State // Prepare-record phase
    \ForAll {group $\gets$ groups} \label{alg:aup:prepare-begin}
      \State records $\gets$ \Call{createPreparedRecords}{group}
      \State // multiple \Call{atomicWrite}{} can be run in parallel
      \State \Call{atomicWrite}{records}
    \EndFor \label{alg:aup:prepare-end}
    \State // Validate-record phase
      \If {\Call{ensureSerializable}{} \textbf{or} \Call{isMetadataDecoupled}{}} \label{alg:aup:validate-begin}
        \ForAll {$(key,record) \gets readSet$}
        \If {\textit{metadata is decoupled}}
          \If {\Call{consistentReadable}{key}}
            \State continue
          \EndIf
         \EndIf
         \State \Call{validateRecord}{$key$, $record$} \Comment{Re-read records}
        \EndFor
    \EndIf \label{alg:aup:validate-end}
      \State // Commit-state phase
      \State \Call{commitState}{TxID} \Comment{Regarded as committed here} \label{alg:aup:commit-state}
      \State // Commit-record phase (asynchronously executable)
    \ForAll {group $\gets$ groups} \label{alg:aup:commit-begin}
      \State records $\gets$ \Call{createCommittedRecords}{group}
      \State // multiple \Call{atomicWrite}{} can be run in parallel
      \State \Call{atomicWrite}{records}
    \EndFor \label{alg:aup:commit-end}
  \EndFunction

  \Function {groupByAtomicityUnit}{writeSet} \label{alg:aup:groupby-aup-begin}
    \State groups $\gets \emptyset$
    \ForAll {w $\gets$ writeSet}
      \State key $\gets \emptyset$
      \State unit $\gets$ \Call{getAtomicityUnit}{w.key} \label{alg:aup:get-atomicity-unit}
      \Switch{unit}
        \Case{\textit{RECORD}}
          \State key.clusteringKey $\gets$ w.clusteringKey
        \EndCase
        \Case{\textit{PARTITION}}
          \State key.partitionKey $\gets$ w.partitionKey
        \EndCase
        \Case{\textit{TABLE}}
          \State key.table $\gets$ w.table
        \EndCase
        \Case{\textit{NAMESPACE}}
          \State key.namespace $\gets$ w.namespace
        \EndCase
        \Case{\textit{STORAGE}}
          \State key.storage $\gets$ w.storage
        \EndCase
      \EndSwitch
      \State \Call{Push}{groups[key], w} \Comment{Group by Atomicity Unit}
    \EndFor
    \State return groups
  \EndFunction \label{alg:aup:groupby-aup-end}
  \State
\end{algorithmic}
\end{algorithm}

\begin{algorithm}[htbp]
\caption{Atomicity Unit Handling in the Database Abstraction Layer}
  \label{alg:atomicity-unit-handling}
  \begin{algorithmic}[1]
  \Function {atomicWrite}{records} \label{alg:auh:atomic-write-begin}
      \State // Identify the database to access
      \State db $\gets$ \Call{getDatabase}{records[0].key}
      \State db.\Call{atomicWrite}{records}
  \EndFunction \label{alg:auh:atomic-write-end}

  \Function {getAtomicityUnit}{key} \label{alg:auh:get-atomicity-unit-begin}
      \State db $\gets$ \Call{getDatabase}{key}
      \State return db.\Call{getAtomicityUnit}{}
  \EndFunction \label{alg:auh:get-atomicity-unit-end}
\end{algorithmic}
\end{algorithm}

\subsubsection{Approach} \label{subsubsec:aup-approach}

AUP is a technique that enables federated transaction management to aggressively push down database operations by making use of the atomicity unit of each underlying database.
By making each database adapter that implements the abstraction declare its atomicity unit, the transaction manager can effectively group writes for each database.
As a result, each underlying database accepts a set of grouped operations in a batch and writes them as a single database transaction, reducing the number of writes to disk and fully harnessing the power and optimization opportunities of the databases.

Specifically, when a federated transaction writes to multiple databases, writes for preparing records for each database and writes for committing records for each database are grouped together, respectively.
When a federated transaction writes to one database, all the writes for preparing records, the write for committing to the coordinator table, and all the writes for committing records are grouped together.
Since all these writes are executed atomically in the underlying database, the writes for preparing records and committing to the coordinator table can be omitted, resulting in only the writes for committing records being issued.
While using federated transactions for a single database may initially seem unnecessary, it can be beneficial during the early phases of projects. For instance, a system may start with one database and one microservice, but as development progresses and more features are added, additional databases and microservices might be created.

\subsubsection{Algorithm} \label{subsubsec:aup-algorithm}

Algorithm \ref{alg:aup} shows the algorithm of AUP.
AUP modifies the behavior of the \textsc{commit} method, called when a transaction is ready to commit.

When a client begins a transaction, it first generates a transaction ID (TxID).
Then, when the client is ready to commit the transaction after performing operations, such as \textit{get}, \textit{insert}, \textit{update}, for reading and writing records, it calls \textsc{commit} (line \ref{alg:aup:commit}) to request ScalarDB to commit the transaction.
Note that ScalarDB uses a single-version optimistic concurrency control; thus, ScalarDB holds the read set (\textit{readSet}) and write set (\textit{writeSet}) of the transaction in its local memory space at the time of committing.
ScalarDB with AUP first groups the records of the write set by their atomicity units (line \ref{alg:aup:groupby-aup}).
We later explain the case where records are committed in one phase (lines \ref{alg:aup:one-phase-begin}-\ref{alg:aup:one-phase-end}).
Then, ScalarDB prepares each group in one shot with an atomic write of the underlying database, instead of preparing the records of the group one by one (lines \ref{alg:aup:prepare-begin}-\ref{alg:aup:prepare-end}).
Note that the atomic write capability of databases is abstracted in the database abstraction layer (lines \ref{alg:auh:atomic-write-begin}-\ref{alg:auh:atomic-write-end} of Algorithm \ref{alg:atomicity-unit-handling}) and is implemented in each adapter.
ScalarDB checks conflicting preparations by using linearizable conditional writes; a transaction updates a record if the record has not been updated by another transaction since the transaction read it by checking if the TxID of the record has not been changed.
If all the preparation succeeds, it commits the transaction by writing a COMMITTED state record to the coordinator table (line \ref{alg:aup:commit-state}).
Note that writing to the coordinator table is also done using linearizable conditional writes to coordinate concurrent writes; creating a state record with a TxID if there is no record for the TxID.
Once the COMMITTED state is properly written to the coordinator table, the transaction is regarded as committed.
Then, ScalarDB commits all the prepared groups by changing the states of the records of the groups to COMMITTED (lines \ref{alg:aup:commit-begin}-\ref{alg:aup:commit-end}).
When the size of the grouped writes is one (lines \ref{alg:aup:one-phase-begin}-\ref{alg:aup:one-phase-end}), AUP executes a transaction in one phase, only writing for committing records, as discussed in Section \ref{subsubsec:aup-approach}.
We skipped the explanation of a validate-record phase (lines \ref{alg:aup:validate-begin}-\ref{alg:aup:validate-end}) to make transactions strict serializable because AUP does not change the logic of the phase.
More details about the validation phase can be found in the ScalarDB paper \cite{yamada-scalardb}.

Let us explain how to group the write set (lines \ref{alg:aup:groupby-aup-begin}-\ref{alg:aup:groupby-aup-end}).
ScalarDB begins by examining each write in the write set to identify its atomicity unit (line \ref{alg:aup:get-atomicity-unit} of Algorithm \ref{alg:aup} and lines \ref{alg:auh:atomic-write-begin}-\ref{alg:auh:atomic-write-end} of Algorithm \ref{alg:atomicity-unit-handling}).
Next, it determines a key called the \textit{group key}, which is used to group records based on their atomicity unit.
Specifically, the group key is derived from the atomicity unit of the database where the corresponding record is stored. For instance, if a record's atomicity unit is RECORD, the group key for the record is a pointer to it, consisting of the primary key (the partition key and the clustering key), the table name, the namespace name, and the database name.\footnote{ScalarDB organizes records into tables, which are contained within namespaces that belong to a database.}
Similarly, if the atomicity unit is TABLE, the group key points to the entire table containing the record, consisting of the table name, the namespace name, and the database name.
This method allows records with the same group key to be grouped together, facilitating atomic writes to the database for each group.

It is important to note that the code does not include any database-specific processing. By defining the atomicity unit within the database abstraction and requiring each adapter to declare its level, the database-agnostic transaction manager can optimize its processing by utilizing database-specific capabilities without breaking the abstraction.

\subsection{AU-Scoped Decoupling}

As discussed in Section \ref{subsec:challenges}, database-agnostic federated transactions require that application data and the corresponding transaction metadata be located within the same record to facilitate efficient updates.
This necessitates a schema migration for running federated transactions on existing databases. 
Managing transaction metadata in separate locations increases the number of read and write operations to the underlying database more than doubles, which can severely impact performance.
To address this issue, we introduce \textit{AU-scoped decoupling (AUD)} for transaction metadata, a technique designed to achieve metadata decoupling while minimizing performance degradation by leveraging AU.

\subsubsection{Approach}

AUD extends database abstraction to efficiently manage decoupled metadata.
AUD virtually integrates application data and the associated metadata, allowing the transaction manager to treat the decoupled data like a regular record that contains both data.

When writing a record, AUD separates it into two parts (application part and metadata part) and writes them individually using the \textsc{atomicWrite} of the database abstraction. Importantly, as long as the decoupled records are located within the atomicity unit of the underlying database, they are grouped together and written as a single transaction of the database.

For reading records, AUD retrieves both the application record and the corresponding metadata record, joining them afterward in the abstraction layer. Since these reads could occur as separate transactions or operations, they may not originate from the same snapshot. Therefore, in such a case, AUD performs additional reads before committing to ensure that the records have not changed since the initial reads, similar to the read-validation process in optimistic concurrency control.
Actually, such additional reads can be done in the same code (lines \ref{alg:aup:validate-begin}-\ref{alg:aup:validate-end}) as the read-validation, as shown in Algorithm \ref{alg:aup}.

\subsubsection{Optimizations}

Joining application data and metadata in the abstraction layer and validating the data in the transaction management layer is a robust way since it works for any databases that meet the requirements of ScalarDB, 
but it is not efficient.
If the application data and the metadata of records are stored in the same database, and the database provides a transaction capability with snapshot isolation or stricter isolation, which we abstract as \textit{consistent readability}, reading both data can be done without such re-reading by using the database's transaction.

Furthermore, if a database provides a database view or equivalent feature, which we abstract as \textit{view joinability}, the join processing can be pushed down to the database layer.

AUD extends database abstraction for consistent readability and view joinability.
By requiring each database adapter to declare its capabilities for those, the database-agnostic transaction manager can optimize the handling of decoupled metadata whenever possible.

\subsubsection{Algorithm}

Algorithm \ref{alg:metadata-decoupling} shows the algorithm of AUD.
It modifies the read and write logic of the database abstraction.

For writing records (lines \ref{alg:aud:atomic-write-begin}-\ref{alg:aud:atomic-write-end}), the \textsc{atomicWrite} method is modified. It first \textit{decouples} the records into application data and the corresponding metadata, identifies a database to access based on the records, and write them all atomically to the database.
Note that the \textsc{atomicWrite} method assumes that the records it accepts are all in the same atomicity unit, so that all the data and metadata of the records are grouped into one transaction of the underlying database.

For reading records (lines \ref{alg:aud:read-begin}-\ref{alg:aud:read-end}), if the application data and the corresponding metadata of records are not \textit{consistently readable},
it reads both data in separate transactions or operations and join them (line \ref{alg:aud:read-decoupled}, lines \ref{alg:aud:read-decoupled-begin}-\ref{alg:aud:read-decoupled-end}).
As discussed, these reads are done again in the validation phase of the transaction in the upper layer to guarantee these see the same snapshot (lines \ref{alg:aup:validate-begin}-\ref{alg:aup:validate-end} of Algorithm \ref{alg:aup}).
If both data are \textit{consistently readable},
reading these data can be done without such re-reading by using the database's transaction (lines \ref{alg:aud:read-decoupled-in-abstraction-begin}-\ref{alg:aud:read-decoupled-in-abstraction-end}) or using the database's view feature (lines \ref{alg:aud:read-decoupled-from-view-begin}-\ref{alg:aud:read-decoupled-from-view-end}).

\begin{algorithm}[htbp]
\caption{AU-Scoped Decoupling for Transaction Metadata}
  \label{alg:metadata-decoupling}
  \begin{algorithmic}[1]
  \Function {atomicWrite}{records} \label{alg:aud:atomic-write-begin}
      \State [app\_records, meta\_records] $\gets$ \Call{decouple}{records}
      \State // Identify the database to access
      \State db $\gets$ \Call{getDatabase}{app\_records[0].key}
      \State db.\Call{atomicWrite}{[app\_records, meta\_records]}
  \EndFunction \label{alg:aud:atomic-write-end}

  \Function {read}{key} \label{alg:aud:read-begin} \Comment{the returned record will be inserted into readSet}
      \If {\textit{metadata is decoupled}}
        \If {\Call{consistentReadable}{key}}
          \If {\Call{viewJoinable}{key}}
            \State return \Call{readDecoupledFromView}{key} \label{alg:aud:read-decoupled-from-view}
          \Else 
            \State return \Call{readDecoupledInAbstraction}{key} \label{alg:aud:read-decoupled-in-abstraction}
          \EndIf
        \Else
          \State return \Call{readDecoupled}{key} \label{alg:aud:read-decoupled}
        \EndIf
      \Else
          \State return \Call{readNormally}{key}
      \EndIf
  \EndFunction \label{alg:aud:read-end}

  \Function {readDecoupledFromView}{key} \label{alg:aud:read-decoupled-from-view-begin}
      \State db $\gets$ \Call{getDatabase}{key}
      \State view $\gets$ db.\Call{getView}{key} \Comment{use the db's view}
      \State record $\gets$ view.\Call{read}{key}
      \State return record
  \EndFunction \label{alg:aud:read-decoupled-from-view-end}

  \Function {readDecoupledInAbstraction}{key} \label{alg:aud:read-decoupled-in-abstraction-begin}
      \State db $\gets$ \Call{getDatabase}{key}
      \State tx $\gets$ db.\Call{begin}{} \Comment{use the db's transaction}
      \State meta $\gets$ tx.\Call{getMetaDatabase}{key}
      \State app\_record $\gets$ tx.\Call{read}{key}
      \State meta\_record $\gets$ meta.\Call{read}{key}
      \State tx.\Call{commit}{}
      \State return \Call{join}{app\_record, meta\_record}
  \EndFunction \label{alg:aud:read-decoupled-in-abstraction-end}

  \Function {readDecoupled}{key} \label{alg:aud:read-decoupled-begin}
      \State db $\gets$ \Call{getDatabase}{key}
      \State metaDb $\gets$ \Call{getMetaDatabase}{key}
      \State app\_record $\gets$ db.\Call{read}{key}
      \State meta\_record $\gets$ metaDb.\Call{read}{key}
      \State return \Call{join}{app\_record, meta\_record}
  \EndFunction \label{alg:aud:read-decoupled-end}

  \Function {readNormally}{key} \label{alg:AUD:read-normally}
      \State db $\gets$ \Call{getDatabase}{key}
      \State return db.\Call{read}{key}
  \EndFunction

  \Function {consistentReadable}{key} \label{alg:AUD:consistent-readable}
      \State unit $\gets$ \Call{getAtomicityUnit}{key}
      \If {\textit{multiple records are consistently readable within the unit}}
        \State return true
      \Else
        \State return false
      \EndIf
  \EndFunction

\end{algorithmic}
\end{algorithm}

\section{Evaluation}

In this section, we evaluate the approaches discussed in Section \ref{sec:au}.

\subsection{Benchmarked Systems}

We implemented the proposed approaches in a database-agnostic transaction manager, ScalarDB \cite{db/scalardb}.
We used ScalarDB 3.17.1.
We enabled all the configurations of parallel execution starting with \texttt{\seqsplit{scalar.db.consensus\_commit.parallel\_*}}.
We also enabled all the configurations of asynchronous execution starting with \texttt{\seqsplit{scalar.db.consensus\_commit.async\_*}}.
We used PostgreSQL version 18.1 for the underlying database of ScalarDB.

\subsection{Workload}

We used YCSB \cite{bench/cooper} to clarify the performance differences of the benchmarked systems.
YCSB is a benchmark commonly used for key-value store evaluation and also adopted in transactional database evaluation by accessing multiple records in a single transaction.
We used scalardb-benchmarks \cite{soft/scalardb-benchmarks} as a benchmarking tool.
We used two types of workloads: Workload-F (read-modify-write workload) and Workload-C (read-only workload), with uniform request distribution.

\begin{figure*}[t]
    \begin{center}
        \subfigure[1 RMW operation per database.]{
            \includegraphics[width=88mm]{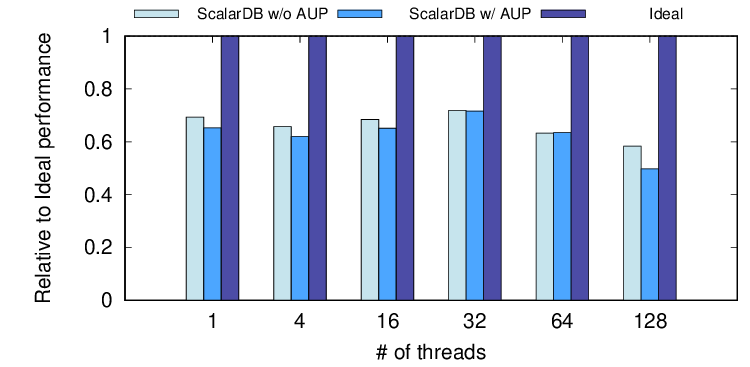}
            \label{fig:aup-1op}}
        \hspace{-10pt}
        \subfigure[4 RMW operations per database.]{
            \includegraphics[width=88mm]{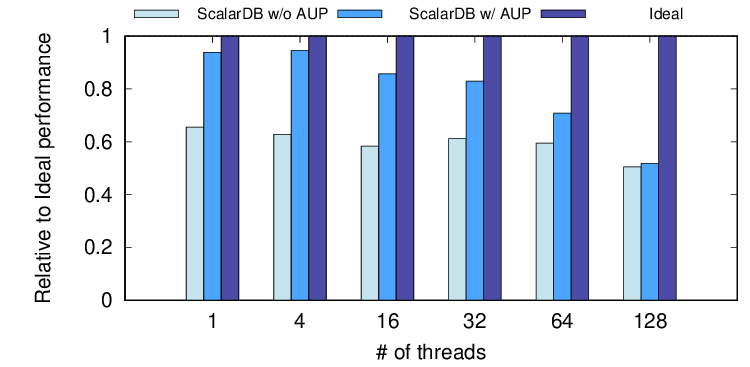}
            \label{fig:aup-4op}}
        \hspace{-10pt}
        \subfigure[8 RMW operations per database.]{
            \includegraphics[width=88mm]{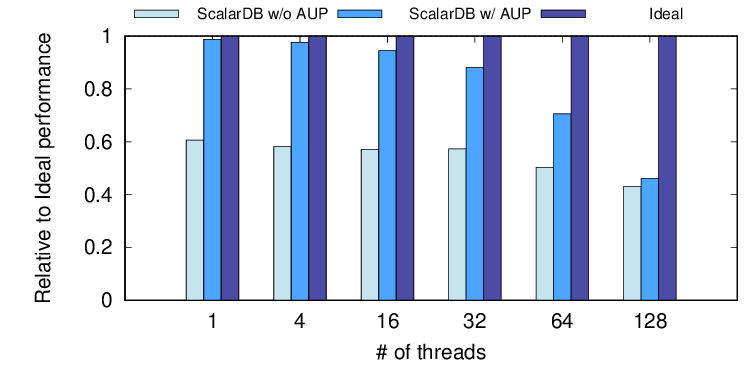}
            \label{fig:aup-8op}}
        \hspace{-10pt}
        \subfigure[16 RMW operations per database.]{
            \includegraphics[width=88mm]{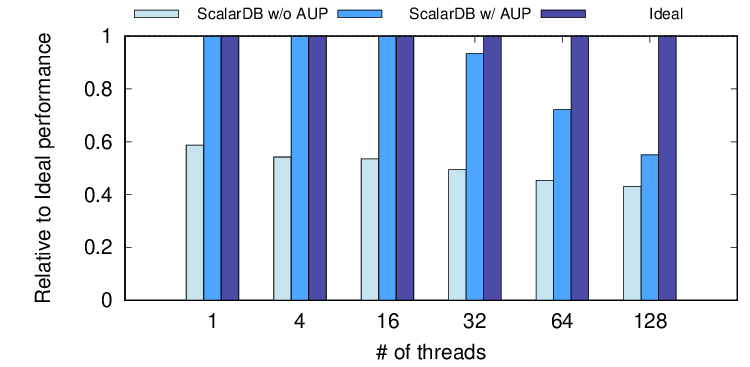}
            \label{fig:aup-16op}}
        \hspace{-10pt}

        \caption{Performance benefit of AUP.}
        \label{fig:aup-evaluation}
    \end{center}
\end{figure*}

\subsection{Experimental Setup}

All experiments were conducted with Microsoft Azure.
For each database instance, we used a Standard\_D8ds\_v4 instance (8 vCores, 32 GiB memory, 12800 max iops).
For the experiments for evaluating AUP, we used two PostgreSQL instances for managing application data and one PostgreSQL instance for the coordinator.
For the experiments for evaluating AUD, we used one PostgreSQL instance for managing application data and the coordinator.
For both experiments, we used a Standard D16s v5 (16 vcpus, 64 GiB memory) instance for clients that issue transactions to databases through ScalarDB.

\begin{figure*}[h]
    \begin{center}
        \subfigure[1 read-modify-write operation (Workload-F).]{
            \includegraphics[width=88mm]{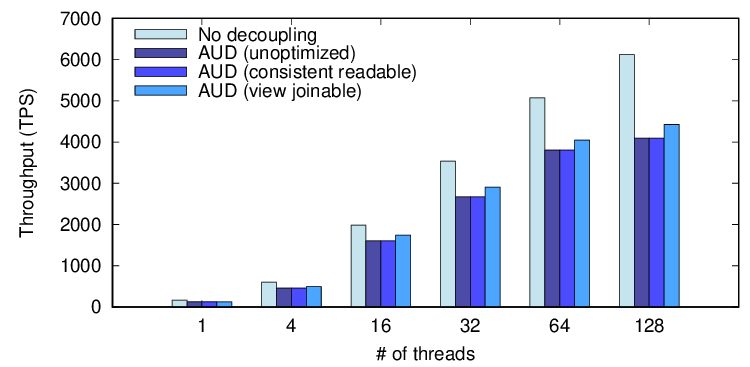}
            \label{fig:aud-ycsbf-128-1op}}
        \hspace{-10pt}
        \subfigure[8 read-modify-write operations (Workload-F).]{
            \includegraphics[width=88mm]{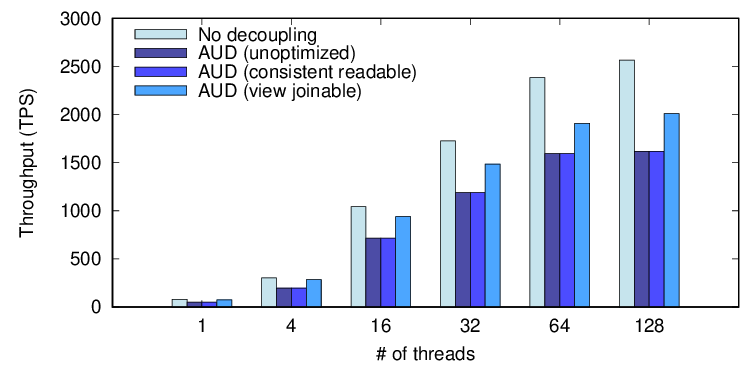}
            \label{fig:aud-ycsbf-128-8op}}
        \hspace{-10pt}
        \subfigure[1 read operation (Workload-C).]{
            \includegraphics[width=88mm]{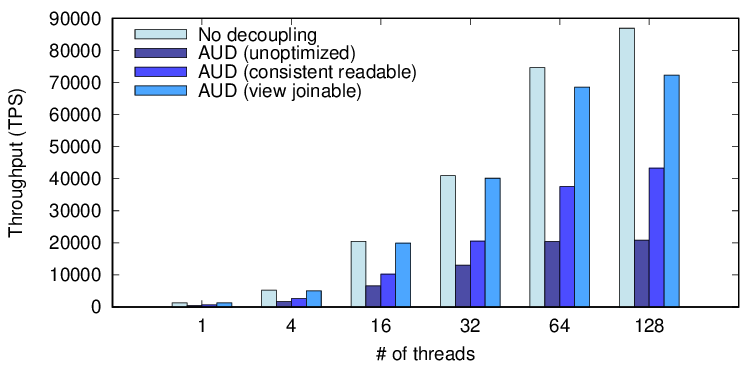}
            \label{fig:aud-ycsbc-128-1op}}
        \hspace{-10pt}
        \subfigure[8 read operations (Workload-C)]{
            \includegraphics[width=88mm]{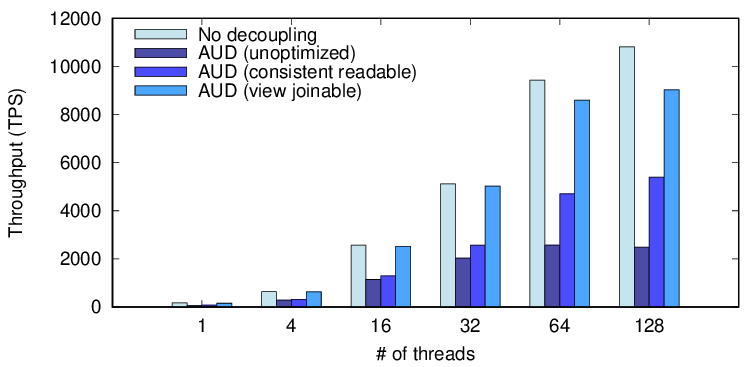}
            \label{fig:aud-ycsbc-128-8op}}
        \hspace{-10pt}

        \caption{Effectiveness of AUD on YCSB.}
        \label{fig:aud-ycsbf-evaluation}
    \end{center}
\end{figure*}

\subsection{Performance benefit of AUP}

This section evaluates how AUP accelerates database-agnostic transactions.
In the experiments, we deployed two PostgreSQL instances for managing application data and one PostgreSQL instance for the transaction coordinator and its data.
We used a modified version of YCSB Workload-F for federated transactions, which issues a specified number of operations to both instances in a transactional manner.
We configured the benchmark to use 128 bytes payload and loaded 1 million records before the experiments.

We compared ScalarDB without AUP (\textbf{ScalarDB w/o AUP}) to ScalarDB with AUP (\textbf{ScalarDB w/ AUP}).
Additionally, we compared both configurations to the ideal performance scenario (\textbf{Ideal}), where clients issue operations to multiple databases directly via JDBC without transaction coordination between them.
It is important to note that \textbf{Ideal} does not provide the same correctness guarantees (ACID) as the others but serves as a benchmark for the best possible performance that can be achieved.

Figure \ref{fig:aup-evaluation} shows the results.
The graph presents the relative throughput compared to the ideal scenario for each number of client threads.
We observed that AUP significantly enhances database-agnostic transaction performance, particularly when the number of operations increases.
This improvement occurs because as the transaction manager handles more operations, more of them can be grouped into a single underlying database transaction.

Let us analyze the performance differences between \textbf{ScalarDB w/ AUP} and \textbf{Ideal} in more detail.
The performance differences are mainly from how they write records when committing transactions since reading records are mostly from underlying databases' buffers.
When \textbf{ScalarDB w/ AUP} commits a transaction, it performs several actions: it writes records atomically with one transaction for each database for preparation, writes a record to the coordinator for committing the transaction, and writes records atomically with one transaction to commit the records. In contrast, the \textbf{Ideal} case involves only the first (or the last) record writing.
The last phase of ScalarDB is executed asynchronously, as configured. This means that when there are not many concurrent transactions, the performance difference primarily arises from the write operations to the coordinator. As a result, in Figures \ref{fig:aup-4op}, \ref{fig:aup-8op}, and \ref{fig:aup-16op}, the performance of \textbf{ScalarDB w/ AUP} is similar to that of the \textbf{Ideal} case.
However, as the number of client threads increases, overall performance is influenced by the number of write operations performed on the underlying databases. Consequently, \textbf{ScalarDB w/ AUP} exhibits performance that is about half that of the \textbf{Ideal} case.

We confirmed that AUP significantly enhances the performance of database-agnostic transactions when the transactions contain multiple operations and the level of concurrency is not too high.

\subsection{Effectiveness of AUD}

This section evaluates how ScalarDB with AUD efficiently handles database-agnostic transactions when the transaction metadata is decoupled.
In the experiments, we deployed one PostgreSQL instance for managing application data and the transaction coordinator and its data.
We used Workload-F and Workload-C.
We configured the benchmark to use 128-byte payload and loaded 1 million records before the experiments.

\textbf{No decoupling} is a case where the transaction metadata is stored in the corresponding application record.
\textbf{AUD (unoptimized)} is a case where the transaction metadata is separately managed in a different table of the same database. We employed AUD to manage the decoupled transaction metadata, but AUD does not utilize the consistent readability of the underlying database.
\textbf{AUD (consistent readable)} instead utilizes the consistent readability of the underlying database.
\textbf{AUD (view joinable)} further utilizes the view joinability of the underlying database.
Note that we disabled the one-phase commit optimization\footnote{We set \texttt{\seqsplit{scalar.db.consensus\_commit.one\_phase\_commit.enabled}} to false.}, where the writes for preparing records and committing to the coordinator table can be omitted as discussed in Section \ref{subsubsec:aup-algorithm}, to clarify the difference of the approaches.

Figure \ref{fig:aud-ycsbf-128-1op} and \ref{fig:aud-ycsbf-128-8op} show the results on Workload-F, configured with a payload size of 128 bytes.
\textbf{No decoupling} performed the best because reads and writes were not amplified.
\textbf{AUD (consistent readable)} decoupled metadata, which resulted in some overhead due to the need to perform twice as many read and write operations in the abstraction layer. However, thanks to AU, two writes and two reads were issued in a single transaction respectively, which mitigates the performance slowdown to about 30\%.
\textbf{AUD (unoptimized)} had the same performance as \textbf{AUD (consistent readable)} because the keys in the read set and the write set were the same in Workload-F and the read validation was not triggered.
Lastly, \textbf{AUD (view joinable)} improved upon the performance of \textbf{AUD (consistent readable)} by facilitating joins in the underlying database.

Figure \ref{fig:aud-ycsbc-128-1op} and \ref{fig:aud-ycsbc-128-8op} show the results on Workload-C, configured with a payload size of 128 bytes.
\textbf{No decoupling} again performed the best because reads were not amplified.
\textbf{AUD (unoptimized)} needed to perform twice as many read operations for each application read, and also needed to perform the read validation, which resulted in 4 times as many read operations as \textbf{No decoupling}.
\textbf{AUD (consistent readable)} was able to avoid the read validation; thus, it achieved about twice the throughput of \textbf{AUD (unoptimized)}.
Lastly, \textbf{AUD (view joinable)} improved upon the performance of \textbf{AUD (consistent readable)} by facilitating joins in the underlying database.

Overall, AUD significantly mitigates the performance slowdown caused by metadata decoupling when used with a database that ensures consistent readability and joinability.

\section{Related Work}
\label{sec:related}

\subsection{Database-Dependent Federated Transactions}
    
Federated transaction management has been studied since the early 1990s.
Breitbart et al.~\cite{mdbs/breitbart} formalized the problem of ensuring global serializability across autonomous databases, and the textbook by \"{O}zsu and Valduriez~\cite{ozsu-distributed} provides a comprehensive treatment of distributed and federated transaction processing.
The X/Open XA standard~\cite{dt/xa} established a widely adopted interface for coordinating distributed transactions using two-phase commit (2PC)~\cite{gray-2pc} across XA-compliant resource managers.
However, XA requires that all participating databases implement the XA interface, which limits its applicability to traditional relational databases.
NoSQL databases, many NewSQL databases, and cloud-managed database services generally do not support XA.

\subsection{Database-Agnostic Transaction Management}

Deuteronomy \cite{related/levandosuki} and Cherry Garcia~\cite{dey-cherrygarcia} pioneered the concept of database-agnostic transactions by managing transactions entirely outside the database.
ScalarDB~\cite{yamada-scalardb} extended these approaches to provide strict serializability, further performance optimizations, and several critical mechanisms for productization.
Epoxy~\cite{kraft-epoxy} provides multi-version concurrency control for cross-database transactions.

These approaches typically treat the underlying database as an opaque store: each read or write is issued as an independent database operation, forgoing opportunities to batch operations that the database could execute atomically.
Also, all of these systems except Deutronomy require transaction metadata to be colocated with application data, necessitating a schema migration to run federated transactions on top of existing databases.
Our proposed techniques based on atomicity unit address these limitations.

\subsection{Distributed Transaction Optimization}

Extensive research has focused on optimizing distributed transactions within a single database system.
Calvin~\cite{thomson-calvin} eliminates distributed coordination by deterministically ordering transactions before execution.
SLOG~\cite{ren-slog} extends this idea to geo-replicated settings by partitioning the input log.
Spanner~\cite{corbett-spanner} uses TrueTime for externally consistent transactions, and CockroachDB~\cite{cockroachdb} introduces parallel commits to reduce commit latency in its 2PC protocol.
LEAP~\cite{lin-leap} proposes a non-2PC approach that piggybacks commit decisions onto data messages, reducing coordination overhead.
RedT~\cite{lu-redt} optimizes read-only transactions in geo-distributed databases by eliminating the need for cross-region coordination.
Fine-grained re-execution~\cite{dong-batched} batches transactions and re-executes only conflicting ones, improving throughput.
Concurrency Control as a Service (CCaaS)~\cite{zhou-ccaas} decouples concurrency control from the database engine and offers it as a service, enabling flexible protocol selection.
HDCC~\cite{hdcc} proposes a hybrid deterministic concurrency control that adaptively combines deterministic and non-deterministic approaches.

These optimizations target homogeneous, single-system environments where the transaction engine has full control over storage, logging, and scheduling.
In contrast, our work targets heterogeneous environments where the transaction engine does not know the details of underlying databases.

\subsection{Polystore and Multi-Model Systems}

Polystore systems such as BigDAWG~\cite{duggan-bigdawg} provide unified query access across heterogeneous databases and data models.
Lu and Holubov\'{a}~\cite{lu-multimodel} survey multi-model databases that natively support multiple data models within a single system.
These systems primarily focus on query federation---translating and routing queries across different engines---rather than on transactional consistency.
BigDAWG, for example, supports cross-database queries through its island and shim architecture but does not provide cross-database ACID transactions.

Our work is orthogonal to query federation.
We focus specifically on how to achieve database-agnostic transactions across heterogeneous databases without sacrificing correctness, performance, or practicality.

\section{Conclusion}
This paper introduced \textit{atomicity unit} (AU), a simple yet effective concept that enriches the database abstraction of database-agnostic federated
transaction managers with information about the scope within which each underlying database can guarantee atomicity.
Building on AU, we proposed two techniques: \textit{atomicity unit pushdown} (AUP), which groups write operations by their atomicity units to reduce the number
of database transactions and harness database-specific optimizations, and \textit{AU-scoped decoupling} (AUD), which enables the efficient separation of
transaction metadata from application data by leveraging the consistent readability and view joinability of underlying databases.
Importantly, both techniques preserve the database-agnostic property of the transaction manager, as AU is declared by each database adapter within the existing
abstraction layer.
We implemented both techniques in ScalarDB, an open-source database-agnostic federated transaction manager.
Our evaluation using YCSB workloads on PostgreSQL demonstrated that AUP significantly improves transaction throughput, approaching ideal (non-transactional)
performance when transactions contain multiple operations per database.
AUD enables federated transactions on existing databases without schema migration, with consistent readability and view joinability limiting the performance
overhead to approximately 30\% on read-modify-write workloads.

\clearpage

\balance
\bibliographystyle{ACM-Reference-Format}
\bibliography{paper}


\begin{thebibliography}{40}


\ifx \showCODEN    \undefined \def \showCODEN     #1{\unskip}     \fi
\ifx \showDOI      \undefined \def \showDOI       #1{#1}\fi
\ifx \showISBNx    \undefined \def \showISBNx     #1{\unskip}     \fi
\ifx \showISBNxiii \undefined \def \showISBNxiii  #1{\unskip}     \fi
\ifx \showISSN     \undefined \def \showISSN      #1{\unskip}     \fi
\ifx \showLCCN     \undefined \def \showLCCN      #1{\unskip}     \fi
\ifx \shownote     \undefined \def \shownote      #1{#1}          \fi
\ifx \showarticletitle \undefined \def \showarticletitle #1{#1}   \fi
\ifx \showURL      \undefined \def \showURL       {\relax}        \fi
\providecommand\bibfield[2]{#2}
\providecommand\bibinfo[2]{#2}
\providecommand\natexlab[1]{#1}
\providecommand\showeprint[2][]{arXiv:#2}

\bibitem[Atomikos(2023)]%
        {related/atomikos}
\bibfield{author}{\bibinfo{person}{Atomikos}.} \bibinfo{year}{2023}\natexlab{}.
\newblock \bibinfo{title}{Atomikos}.
\newblock \bibinfo{howpublished}{\url{https://www.atomikos.com/Main/WebHome}}.
\newblock


\bibitem[Breitbart et~al\mbox{.}(1992)]%
        {mdbs/breitbart}
\bibfield{author}{\bibinfo{person}{Y. Breitbart}, \bibinfo{person}{H.
  Garcia-Molina}, {and} \bibinfo{person}{A. Silberschatz}.}
  \bibinfo{year}{1992}\natexlab{}.
\newblock \showarticletitle{Overview of Multidatabase Transaction Management}.
\newblock \bibinfo{journal}{\emph{VLDBJ}} \bibinfo{volume}{1},
  \bibinfo{number}{2} (\bibinfo{year}{1992}), \bibinfo{pages}{181--239}.
\newblock


\bibitem[Breitbart and Silberschatz(1988)]%
        {breitbart1988multidatabase}
\bibfield{author}{\bibinfo{person}{Yuri Breitbart} {and} \bibinfo{person}{Avi
  Silberschatz}.} \bibinfo{year}{1988}\natexlab{}.
\newblock \showarticletitle{Multidatabase update issues}. In
  \bibinfo{booktitle}{\emph{Proceedings of the 1988 ACM SIGMOD international
  Conference on Management of Data}}. \bibinfo{pages}{135--142}.
\newblock


\bibitem[Cerny et~al\mbox{.}(2018)]%
        {intro/cerny}
\bibfield{author}{\bibinfo{person}{T. Cerny}, \bibinfo{person}{M. Donahoo},
  {and} \bibinfo{person}{M. Trnka}.} \bibinfo{year}{2018}\natexlab{}.
\newblock \showarticletitle{Contextual Understanding of Microservice
  Architecture: Current and Future Directions}.
\newblock \bibinfo{journal}{\emph{SIGAPP Appl. Comput. Rev.}}
  \bibinfo{volume}{17}, \bibinfo{number}{4} (\bibinfo{year}{2018}),
  \bibinfo{pages}{29–45}.
\newblock
\showISSN{1559-6915}


\bibitem[Chang et~al\mbox{.}(2008)]%
        {related/chang2}
\bibfield{author}{\bibinfo{person}{F. Chang}, \bibinfo{person}{J. Dean},
  \bibinfo{person}{S. Ghemawat}, \bibinfo{person}{W. Hsieh},
  \bibinfo{person}{D. Wallach}, \bibinfo{person}{M. Burrows},
  \bibinfo{person}{T. Chandra}, \bibinfo{person}{A. Fikes}, {and}
  \bibinfo{person}{R. Gruber}.} \bibinfo{year}{2008}\natexlab{}.
\newblock \showarticletitle{Bigtable: A Distributed Storage System for
  Structured Data}.
\newblock \bibinfo{journal}{\emph{ACM Trans. Comput. Syst.}}
  \bibinfo{volume}{26}, \bibinfo{number}{2}, Article \bibinfo{articleno}{4}
  (\bibinfo{year}{2008}), \bibinfo{numpages}{26}~pages.
\newblock


\bibitem[{Cockroach Labs}(2019)]%
        {cockroachdb}
\bibfield{author}{\bibinfo{person}{{Cockroach Labs}}.}
  \bibinfo{year}{2019}\natexlab{}.
\newblock \bibinfo{title}{Parallel Commits: An Atomic Commit Protocol for
  Globally Distributed Transactions}.
\newblock
  \bibinfo{howpublished}{\url{https://www.cockroachlabs.com/blog/parallel-commits/}}.
\newblock


\bibitem[Cooper et~al\mbox{.}(2010)]%
        {bench/cooper}
\bibfield{author}{\bibinfo{person}{B. Cooper}, \bibinfo{person}{A.
  Silberstein}, \bibinfo{person}{E. Tam}, \bibinfo{person}{R. Ramakrishnan},
  {and} \bibinfo{person}{R. Sears}.} \bibinfo{year}{2010}\natexlab{}.
\newblock \showarticletitle{Benchmarking Cloud Serving Systems with YCSB}. In
  \bibinfo{booktitle}{\emph{SoCC}}. \bibinfo{pages}{143–154}.
\newblock


\bibitem[Corbett et~al\mbox{.}(2012)]%
        {corbett-spanner}
\bibfield{author}{\bibinfo{person}{James~C. Corbett}, \bibinfo{person}{Jeffrey
  Dean}, \bibinfo{person}{Michael Epstein}, \bibinfo{person}{Andrew Fikes},
  \bibinfo{person}{Christopher Frost}, \bibinfo{person}{J.~J. Furman},
  \bibinfo{person}{Sanjay Ghemawat}, \bibinfo{person}{Andrey Gubarev},
  \bibinfo{person}{Christopher Heiser}, \bibinfo{person}{Peter Hochschild},
  \bibinfo{person}{Wilson Hsieh}, \bibinfo{person}{Sebastian Kanthak},
  \bibinfo{person}{Eugene Kogan}, \bibinfo{person}{Hongyi Li},
  \bibinfo{person}{Alexander Lloyd}, \bibinfo{person}{Sergey Melnik},
  \bibinfo{person}{David Mwaura}, \bibinfo{person}{David Nagle},
  \bibinfo{person}{Sean Quinlan}, \bibinfo{person}{Rajesh Rao},
  \bibinfo{person}{Lindsay Rolig}, \bibinfo{person}{Yasushi Saito},
  \bibinfo{person}{Michal Szymaniak}, \bibinfo{person}{Christopher Taylor},
  \bibinfo{person}{Ruth Wang}, {and} \bibinfo{person}{Dale Woodford}.}
  \bibinfo{year}{2012}\natexlab{}.
\newblock \showarticletitle{Spanner: {Google}'s Globally Distributed Database}.
  In \bibinfo{booktitle}{\emph{Proceedings of the 10th USENIX Symposium on
  Operating Systems Design and Implementation (OSDI)}}.
  \bibinfo{pages}{261--264}.
\newblock


\bibitem[Dehghani(2020)]%
        {intro/datamesh}
\bibfield{author}{\bibinfo{person}{Z. Dehghani}.}
  \bibinfo{year}{2020}\natexlab{}.
\newblock \bibinfo{title}{Data Mesh Principles and Logical Architecture}.
\newblock
  \bibinfo{howpublished}{\url{https://martinfowler.com/articles/data-mesh-principles.html}}.
\newblock


\bibitem[Dey et~al\mbox{.}(2013)]%
        {dey-cherrygarcia}
\bibfield{author}{\bibinfo{person}{Anamika Dey}, \bibinfo{person}{Alan Fekete},
  {and} \bibinfo{person}{Uwe R\"{o}hm}.} \bibinfo{year}{2013}\natexlab{}.
\newblock \showarticletitle{Scalable Transactions across Heterogeneous {NoSQL}
  Key-Value Data Stores}.
\newblock \bibinfo{journal}{\emph{Proceedings of the VLDB Endowment}}
  \bibinfo{volume}{6}, \bibinfo{number}{12}, \bibinfo{pages}{1434--1439}.
\newblock
\urldef\tempurl%
\url{https://doi.org/10.14778/2536274.2536331}
\showDOI{\tempurl}


\bibitem[Dong et~al\mbox{.}(2023)]%
        {dong-batched}
\bibfield{author}{\bibinfo{person}{Zhihan Dong}, \bibinfo{person}{Haowen Li},
  \bibinfo{person}{Junchang Wang}, {and} \bibinfo{person}{Shuai Mu}.}
  \bibinfo{year}{2023}\natexlab{}.
\newblock \showarticletitle{Fine-Grained Re-Execution for Efficient Batched
  Commit of Distributed Transactions}.
\newblock \bibinfo{journal}{\emph{Proceedings of the VLDB Endowment}}
  \bibinfo{volume}{16}, \bibinfo{number}{8} (\bibinfo{year}{2023}),
  \bibinfo{pages}{1930--1942}.
\newblock
\urldef\tempurl%
\url{https://doi.org/10.14778/3594512.3594521}
\showDOI{\tempurl}


\bibitem[Duggan et~al\mbox{.}(2015)]%
        {duggan-bigdawg}
\bibfield{author}{\bibinfo{person}{Jennie Duggan}, \bibinfo{person}{Aaron~J.
  Elmore}, \bibinfo{person}{Michael Stonebraker}, \bibinfo{person}{Magda
  Balazinska}, \bibinfo{person}{Bill Howe}, \bibinfo{person}{Jeremy Kepner},
  \bibinfo{person}{Sam Madden}, \bibinfo{person}{David Maier},
  \bibinfo{person}{Tim Mattson}, {and} \bibinfo{person}{Stan Zdonik}.}
  \bibinfo{year}{2015}\natexlab{}.
\newblock \showarticletitle{The {BigDAWG} Polystore System}.
\newblock \bibinfo{journal}{\emph{SIGMOD Record}} \bibinfo{volume}{44},
  \bibinfo{number}{2} (\bibinfo{year}{2015}), \bibinfo{pages}{11--16}.
\newblock
\urldef\tempurl%
\url{https://doi.org/10.1145/2814710.2814713}
\showDOI{\tempurl}


\bibitem[Elmagarmid and Du(1990)]%
        {mdbs/elmagarmid}
\bibfield{author}{\bibinfo{person}{A.K. Elmagarmid} {and} \bibinfo{person}{W.
  Du}.} \bibinfo{year}{1990}\natexlab{}.
\newblock \showarticletitle{A paradigm for concurrency control in heterogeneous
  distributed database systems}. In \bibinfo{booktitle}{\emph{[1990]
  Proceedings. Sixth International Conference on Data Engineering}}.
  \bibinfo{pages}{37--46}.
\newblock
\urldef\tempurl%
\url{https://doi.org/10.1109/ICDE.1990.113452}
\showDOI{\tempurl}


\bibitem[Georgakopoulos et~al\mbox{.}(1991)]%
        {mdbs/geo}
\bibfield{author}{\bibinfo{person}{D. Georgakopoulos}, \bibinfo{person}{M.
  Rusinkiewicz}, {and} \bibinfo{person}{A. Sheth}.}
  \bibinfo{year}{1991}\natexlab{}.
\newblock \showarticletitle{On serializability of multidatabase transactions
  through forced local conflicts}. In \bibinfo{booktitle}{\emph{ICDE}}.
  \bibinfo{pages}{314--323}.
\newblock


\bibitem[Goedegebuure et~al\mbox{.}(2024)]%
        {datamesh/goedegebuure}
\bibfield{author}{\bibinfo{person}{Abel Goedegebuure}, \bibinfo{person}{Indika
  Kumara}, \bibinfo{person}{Stefan Driessen}, \bibinfo{person}{Willem-Jan Van
  Den~Heuvel}, \bibinfo{person}{Geert Monsieur}, \bibinfo{person}{Damian~Andrew
  Tamburri}, {and} \bibinfo{person}{Dario~Di Nucci}.}
  \bibinfo{year}{2024}\natexlab{}.
\newblock \showarticletitle{Data Mesh: A Systematic Gray Literature Review}.
\newblock \bibinfo{journal}{\emph{ACM Comput. Surv.}} \bibinfo{volume}{57},
  \bibinfo{number}{1}, Article \bibinfo{articleno}{11} (\bibinfo{date}{Oct.}
  \bibinfo{year}{2024}), \bibinfo{numpages}{36}~pages.
\newblock
\urldef\tempurl%
\url{https://doi.org/10.1145/3687301}
\showDOI{\tempurl}


\bibitem[Gray(1978)]%
        {gray-2pc}
\bibfield{author}{\bibinfo{person}{Jim Gray}.} \bibinfo{year}{1978}\natexlab{}.
\newblock \showarticletitle{Notes on Data Base Operating Systems}. In
  \bibinfo{booktitle}{\emph{Operating Systems: An Advanced Course}}
  \emph{(\bibinfo{series}{Lecture Notes in Computer Science},
  Vol.~\bibinfo{volume}{60})}. \bibinfo{publisher}{Springer},
  \bibinfo{pages}{393--481}.
\newblock
\urldef\tempurl%
\url{https://doi.org/10.1007/3-540-08755-9_9}
\showDOI{\tempurl}


\bibitem[Group(1991)]%
        {dt/xa}
\bibfield{author}{\bibinfo{person}{The~Open Group}.}
  \bibinfo{year}{1991}\natexlab{}.
\newblock \bibinfo{title}{Distributed Transaction Processing: The XA
  Specification}.
\newblock
  \bibinfo{howpublished}{\url{https://pubs.opengroup.org/onlinepubs/009680699/toc.pdf}}.
\newblock


\bibitem[Hong et~al\mbox{.}(2025)]%
        {hdcc}
\bibfield{author}{\bibinfo{person}{Yinhao Hong}, \bibinfo{person}{Hongyao
  Zhao}, \bibinfo{person}{Wei Lu}, \bibinfo{person}{Xiaoyong Du},
  \bibinfo{person}{Yuxing Chen}, \bibinfo{person}{Anqun Pan}, {and}
  \bibinfo{person}{Lixiong Zheng}.} \bibinfo{year}{2025}\natexlab{}.
\newblock \showarticletitle{A Hybrid Approach to Integrating Deterministic and
  Non-Deterministic Concurrency Control in Database Systems}.
\newblock \bibinfo{journal}{\emph{Proc. VLDB Endow.}} \bibinfo{volume}{18},
  \bibinfo{number}{5} (\bibinfo{date}{Jan.} \bibinfo{year}{2025}),
  \bibinfo{pages}{1376–1389}.
\newblock
\showISSN{2150-8097}
\urldef\tempurl%
\url{https://doi.org/10.14778/3718057.3718066}
\showDOI{\tempurl}


\bibitem[J.~Lewis(2014)]%
        {intro/microservices}
\bibfield{author}{\bibinfo{person}{M.~Fowler J.~Lewis}.}
  \bibinfo{year}{2014}\natexlab{}.
\newblock \bibinfo{title}{Microservices}.
\newblock
  \bibinfo{howpublished}{\url{https://martinfowler.com/articles/microservices.html}}.
\newblock


\bibitem[Kraft et~al\mbox{.}(2023)]%
        {kraft-epoxy}
\bibfield{author}{\bibinfo{person}{Peter Kraft}, \bibinfo{person}{Qian Li},
  \bibinfo{person}{Xinjing Zhou}, \bibinfo{person}{Peter Bailis},
  \bibinfo{person}{Michael Stonebraker}, \bibinfo{person}{Xiangyao Yu}, {and}
  \bibinfo{person}{Matei Zaharia}.} \bibinfo{year}{2023}\natexlab{}.
\newblock \showarticletitle{Epoxy: {ACID} Transactions Across Diverse Data
  Stores}.
\newblock \bibinfo{journal}{\emph{Proceedings of the VLDB Endowment}}
  \bibinfo{volume}{16}, \bibinfo{number}{11} (\bibinfo{year}{2023}),
  \bibinfo{pages}{2742--2754}.
\newblock
\urldef\tempurl%
\url{https://doi.org/10.14778/3611479.3611484}
\showDOI{\tempurl}


\bibitem[Laigner et~al\mbox{.}(2021)]%
        {microservices/laigner}
\bibfield{author}{\bibinfo{person}{R. Laigner}, \bibinfo{person}{Y. Zhou},
  \bibinfo{person}{M. Salles}, \bibinfo{person}{Y. Liu}, {and}
  \bibinfo{person}{M. Kalinowski}.} \bibinfo{year}{2021}\natexlab{}.
\newblock \showarticletitle{Data Management in Microservices: State of the
  Practice, Challenges, and Research Directions}.
\newblock \bibinfo{journal}{\emph{PVLDB}} \bibinfo{volume}{14},
  \bibinfo{number}{13} (\bibinfo{year}{2021}), \bibinfo{pages}{3348–3361}.
\newblock
\showISSN{2150-8097}


\bibitem[Levandoski et~al\mbox{.}(2011)]%
        {related/levandosuki}
\bibfield{author}{\bibinfo{person}{J. Levandoski}, \bibinfo{person}{D. Lomet},
  \bibinfo{person}{M. Mokbel}, {and} \bibinfo{person}{K Zhao}.}
  \bibinfo{year}{2011}\natexlab{}.
\newblock \showarticletitle{Deuteronomy: Transaction Support for Cloud Data}.
  In \bibinfo{booktitle}{\emph{CIDR}}.
\newblock


\bibitem[Lin et~al\mbox{.}(2016)]%
        {lin-leap}
\bibfield{author}{\bibinfo{person}{Qian Lin}, \bibinfo{person}{Pengfei Chang},
  \bibinfo{person}{Gang Chen}, \bibinfo{person}{Beng~Chin Ooi},
  \bibinfo{person}{Kian-Lee Tan}, {and} \bibinfo{person}{Zhengkui Wang}.}
  \bibinfo{year}{2016}\natexlab{}.
\newblock \showarticletitle{Towards a Non-{2PC} Transaction Management in
  Distributed Database Systems}. In \bibinfo{booktitle}{\emph{Proceedings of
  the 2016 ACM SIGMOD International Conference on Management of Data}}.
  \bibinfo{pages}{1659--1674}.
\newblock
\urldef\tempurl%
\url{https://doi.org/10.1145/2882903.2882923}
\showDOI{\tempurl}


\bibitem[Lu and Holubov\'{a}(2019)]%
        {lu-multimodel}
\bibfield{author}{\bibinfo{person}{Jiaheng Lu} {and} \bibinfo{person}{Irena
  Holubov\'{a}}.} \bibinfo{year}{2019}\natexlab{}.
\newblock \showarticletitle{Multi-model Databases: A New Journey to Handle the
  Variety of Data}.
\newblock \bibinfo{journal}{\emph{Comput. Surveys}} \bibinfo{volume}{52},
  \bibinfo{number}{3}, Article \bibinfo{articleno}{55} (\bibinfo{year}{2019}),
  \bibinfo{numpages}{38}~pages.
\newblock
\urldef\tempurl%
\url{https://doi.org/10.1145/3323214}
\showDOI{\tempurl}


\bibitem[Lu et~al\mbox{.}(2023)]%
        {lu-redt}
\bibfield{author}{\bibinfo{person}{Shuai Lu}, \bibinfo{person}{Shuai Mu},
  \bibinfo{person}{Gang Chen}, \bibinfo{person}{Luo Chen},
  \bibinfo{person}{Yanzhao Qian}, {and} \bibinfo{person}{Haibo Chen}.}
  \bibinfo{year}{2023}\natexlab{}.
\newblock \showarticletitle{Efficient Distributed Transaction Processing in
  Heterogeneous Networks}.
\newblock \bibinfo{journal}{\emph{Proceedings of the VLDB Endowment}}
  \bibinfo{volume}{16}, \bibinfo{number}{6} (\bibinfo{year}{2023}),
  \bibinfo{pages}{1372--1385}.
\newblock
\urldef\tempurl%
\url{https://doi.org/10.14778/3583140.3583153}
\showDOI{\tempurl}


\bibitem[Mehrotra et~al\mbox{.}(1992a)]%
        {mdbs/mehrotra}
\bibfield{author}{\bibinfo{person}{Sharad Mehrotra}, \bibinfo{person}{Rajeev
  Rastogi}, \bibinfo{person}{Yuri Breitbart}, \bibinfo{person}{Henry~F. Korth},
  {and} \bibinfo{person}{Avi Silberschatz}.} \bibinfo{year}{1992}\natexlab{a}.
\newblock \showarticletitle{The concurrency control problem in multidatabases:
  characteristics and solutions}. In \bibinfo{booktitle}{\emph{Proceedings of
  the 1992 ACM SIGMOD International Conference on Management of Data}} (San
  Diego, California, USA) \emph{(\bibinfo{series}{SIGMOD '92})}.
  \bibinfo{publisher}{Association for Computing Machinery},
  \bibinfo{address}{New York, NY, USA}, \bibinfo{pages}{288–297}.
\newblock
\showISBNx{0897915216}
\urldef\tempurl%
\url{https://doi.org/10.1145/130283.130327}
\showDOI{\tempurl}


\bibitem[Mehrotra et~al\mbox{.}(1992b)]%
        {mdbs/mehrotra2}
\bibfield{author}{\bibinfo{person}{S. Mehrotra}, \bibinfo{person}{R. Rastogi},
  \bibinfo{person}{A. Silberschatz}, {and} \bibinfo{person}{H. Korth}.}
  \bibinfo{year}{1992}\natexlab{b}.
\newblock \showarticletitle{A Transaction Model for Multidatabase Systems}. In
  \bibinfo{booktitle}{\emph{ICDCS}}. \bibinfo{pages}{56--63}.
\newblock


\bibitem[Newman(2021)]%
        {microservices/newman}
\bibfield{author}{\bibinfo{person}{Sam Newman}.}
  \bibinfo{year}{2021}\natexlab{}.
\newblock \bibinfo{booktitle}{\emph{Building Microservices}
  (\bibinfo{edition}{2nd} ed.)}.
\newblock \bibinfo{publisher}{O'Reilly Media, Inc.}
\newblock
\showISBNx{9781492034025}


\bibitem[Oracle(2023)]%
        {related/tuxedo}
\bibfield{author}{\bibinfo{person}{Oracle}.} \bibinfo{year}{2023}\natexlab{}.
\newblock \bibinfo{title}{Oracle Tuxedo}.
\newblock
  \bibinfo{howpublished}{\url{https://www.oracle.com/middleware/technologies/tuxedo.html}}.
\newblock


\bibitem[\"{O}zsu and Valduriez(2020)]%
        {ozsu-distributed}
\bibfield{author}{\bibinfo{person}{M.~Tamer \"{O}zsu} {and}
  \bibinfo{person}{Patrick Valduriez}.} \bibinfo{year}{2020}\natexlab{}.
\newblock \bibinfo{booktitle}{\emph{Principles of Distributed Database Systems}
  (\bibinfo{edition}{4th} ed.)}.
\newblock \bibinfo{publisher}{Springer}.
\newblock
\urldef\tempurl%
\url{https://doi.org/10.1007/978-3-030-26253-2}
\showDOI{\tempurl}


\bibitem[Ren et~al\mbox{.}(2019)]%
        {ren-slog}
\bibfield{author}{\bibinfo{person}{Kun Ren}, \bibinfo{person}{Dennis Li}, {and}
  \bibinfo{person}{Daniel~J. Abadi}.} \bibinfo{year}{2019}\natexlab{}.
\newblock \showarticletitle{{SLOG}: Serializable, Low-Latency, Geo-Replicated
  Transactions}.
\newblock \bibinfo{journal}{\emph{Proceedings of the VLDB Endowment}}
  \bibinfo{volume}{12}, \bibinfo{number}{11} (\bibinfo{year}{2019}),
  \bibinfo{pages}{1747--1761}.
\newblock
\urldef\tempurl%
\url{https://doi.org/10.14778/3342263.3342647}
\showDOI{\tempurl}


\bibitem[Scalar(2026a)]%
        {db/scalardb}
\bibfield{author}{\bibinfo{person}{Scalar}.} \bibinfo{year}{2026}\natexlab{a}.
\newblock \bibinfo{title}{ScalarDB}.
\newblock
  \bibinfo{howpublished}{\url{https://github.com/scalar-labs/scalardb}}.
\newblock


\bibitem[Scalar(2026b)]%
        {soft/scalardb-benchmarks}
\bibfield{author}{\bibinfo{person}{Scalar}.} \bibinfo{year}{2026}\natexlab{b}.
\newblock \bibinfo{title}{ScalarDB Benchmarks}.
\newblock
  \bibinfo{howpublished}{\url{https://github.com/scalar-labs/scalardb-benchmarks}}.
\newblock


\bibitem[Seata(2023)]%
        {related/seata}
\bibfield{author}{\bibinfo{person}{Seata}.} \bibinfo{year}{2023}\natexlab{}.
\newblock \bibinfo{title}{Seata}.
\newblock \bibinfo{howpublished}{\url{https://seata.io/en-us/}}.
\newblock


\bibitem[Sheth and Larson(1990)]%
        {intro/sheth}
\bibfield{author}{\bibinfo{person}{A. Sheth} {and} \bibinfo{person}{J.
  Larson}.} \bibinfo{year}{1990}\natexlab{}.
\newblock \showarticletitle{Federated Database Systems for Managing
  Distributed, Heterogeneous, and Autonomous Databases}.
\newblock \bibinfo{journal}{\emph{ACM Comput. Surv.}} \bibinfo{volume}{22},
  \bibinfo{number}{3} (\bibinfo{year}{1990}), \bibinfo{pages}{183–236}.
\newblock


\bibitem[Stonebraker and Cetintemel(2005)]%
        {intro/stonebraker}
\bibfield{author}{\bibinfo{person}{M. Stonebraker} {and} \bibinfo{person}{U.
  Cetintemel}.} \bibinfo{year}{2005}\natexlab{}.
\newblock \showarticletitle{"One Size Fits All": An Idea Whose Time Has Come
  and Gone}. In \bibinfo{booktitle}{\emph{ICDE}}. \bibinfo{pages}{2–11}.
\newblock


\bibitem[Tang et~al\mbox{.}(2025)]%
        {mdbs/tang}
\bibfield{author}{\bibinfo{person}{Chuzhe Tang}, \bibinfo{person}{Zhaoguo
  Wang}, \bibinfo{person}{Jinyang Li}, {and} \bibinfo{person}{Haibo Chen}.}
  \bibinfo{year}{2025}\natexlab{}.
\newblock \showarticletitle{Sonata: Multi-Database Transactions Made Fast and
  Serializable}.
\newblock \bibinfo{journal}{\emph{Proc. VLDB Endow.}} \bibinfo{volume}{18},
  \bibinfo{number}{10} (\bibinfo{date}{June} \bibinfo{year}{2025}),
  \bibinfo{pages}{3449–3462}.
\newblock
\showISSN{2150-8097}
\urldef\tempurl%
\url{https://doi.org/10.14778/3748191.3748207}
\showDOI{\tempurl}


\bibitem[Thomson et~al\mbox{.}(2012)]%
        {thomson-calvin}
\bibfield{author}{\bibinfo{person}{Alexander Thomson},
  \bibinfo{person}{Thaddeus Diamond}, \bibinfo{person}{Shu-Chun Weng},
  \bibinfo{person}{Kun Ren}, \bibinfo{person}{Philip Shao}, {and}
  \bibinfo{person}{Daniel~J. Abadi}.} \bibinfo{year}{2012}\natexlab{}.
\newblock \showarticletitle{Calvin: Fast Distributed Transactions for
  Partitioned Database Systems}. In \bibinfo{booktitle}{\emph{Proceedings of
  the 2012 ACM SIGMOD International Conference on Management of Data}}.
  \bibinfo{pages}{1--12}.
\newblock
\urldef\tempurl%
\url{https://doi.org/10.1145/2213836.2213838}
\showDOI{\tempurl}


\bibitem[Yamada et~al\mbox{.}(2023)]%
        {yamada-scalardb}
\bibfield{author}{\bibinfo{person}{Hiroyuki Yamada}, \bibinfo{person}{Toshihiro
  Suzuki}, \bibinfo{person}{Yuji Ito}, {and} \bibinfo{person}{Jun Nemoto}.}
  \bibinfo{year}{2023}\natexlab{}.
\newblock \showarticletitle{{ScalarDB}: Universal Transaction Manager for
  Polystores}.
\newblock \bibinfo{journal}{\emph{Proceedings of the VLDB Endowment}}
  \bibinfo{volume}{16}, \bibinfo{number}{12} (\bibinfo{year}{2023}),
  \bibinfo{pages}{3768--3780}.
\newblock
\urldef\tempurl%
\url{https://doi.org/10.14778/3611540.3611563}
\showDOI{\tempurl}


\bibitem[Zhou et~al\mbox{.}(2025)]%
        {zhou-ccaas}
\bibfield{author}{\bibinfo{person}{Weixing Zhou} {et~al\mbox{.}}}
  \bibinfo{year}{2025}\natexlab{}.
\newblock \showarticletitle{Concurrency Control as a Service}.
\newblock \bibinfo{journal}{\emph{Proceedings of the VLDB Endowment}}
  \bibinfo{volume}{18}, \bibinfo{number}{11} (\bibinfo{year}{2025}),
  \bibinfo{pages}{2761--2774}.
\newblock


\end{thebibliography}

\end{document}